\documentclass[aps,pra,twocolumn,groupedaddress,amsmath,amssymb,floatfix]{revtex4-1}
\usepackage{amsmath,graphicx,dcolumn,epsfig,bm}
\begin{document}

\title{Pressure broadening and shift of D1 line of Ag by He, Ar and N$_{2}$}

\author{T.~Karaulanov }\email[]{kulan@lanl.gov}
\affiliation{Los Alamos National Laboratory, P21 Group, MS D454, Los Alamos, NM 87544}

\author{B.~K.~Park }
\author{D.~Budker}
\affiliation{Department of Physics, University of California at
Berkeley, Berkeley, California 94720-7300}

\author{A.~O.~Sushkov}
\affiliation{Department of Physics, Harvard University, Cambridge, Massachusetts 02138, USA}

\date{\today}

\begin{abstract}
We have studied experimentally pressure broadening and shift of Ag D1 line caused by He, Ar and N$_{2}$ buffer gases. The measurements were done in a heat-pipe type absorption cell at a temperature of $\approx$1000 K and gas pressures up to 1000 torr. The measured values for pressure broadening and shift (in MHz/Torr) are as follows: Ag--He 5.8(1), +1.16(2); Ag--Ar 5.2(2), -2.28(4); Ag--N$_{2}$ 5.2(1), -2.52(7). The ``+'' and ``--'' signs indicate the direction for the shifts to the blue and red side of the spectrum, respectively.
\end{abstract}


%32.70.Jz Line shapes, widths, and shifts
%34. Atomic and molecular collision processes and interactions

\maketitle

\section{\label{Sec:Intro}Introduction}
Broadening and shift of atomic spectral lines caused by buffer gases provide a valuable information about interaction potentials. Studies of these effects in atoms with a single valence electron have focused mainly on alkali atoms, stimulated by the development of atomic clocks, magnetometers, and gyroscopes, as well as by the work on spin-exchange optical pumping important for practical applications. Recently, silver (Ag) attracted attention with its ultra-narrow (1 Hz) optical clock transition between the ground state and the metastable $4d^{9}5s\,^2D_{5/2}$ state \cite{AgClock1, AgClock2}. The study of Ag-He exciplexes \cite{AgExcip} at 5-25~K allowed the reconstruction of Ag-He pair potentials for the $^2$P$_{1/2}$, $^2$P$_{3/2}$, and $^2$D$_{5/2}$ Ag states. Magneto-optical trapping of silver was realized by Brahmss et al.~\cite{AgMOT}. Recently, magnetic trapping of Ag in He buffer gas in the temperature range  of 150~mK--1000~mK was achieved \cite{AgMagT} showing an anomalous temperature dependence of spin relaxation in the Ag-He system that is inconsistent with the model of spin-rotation coupling.

Despite the recent surge of interest in Ag, we found only a single paper, \cite{AgBShChen} discussing pressure broadening and shift of the Ag D1 and D2 spectral lines by Ar and He. The work \cite{AgBShChen} was completed in the 1950s; a spectrograph was used as the spectrum analyzer and a hydrogen discharge tube as the light source. Low spectral resolution limited the lower end of pressures of the buffer gases where pressure effects could be observed to about 6 atm.

In the present work, we measured pressure broadening and shift of D1 Ag line caused by Ar, He and N$_{2}$ using a tunable laser source and covering the pressure range from vacuum to 1000~torr of the buffer gas.

Our work is motivated, on the one hand, by the ongoing study of silver atoms confined in a cryogenic buffer gas \cite{AgCryoLife}, and on the other, possibility of using silver as a proxy for the alkali atoms, primarily sodium, important for the laser guide star applications \cite{SodiumGS} and, more generally, for understanding of the atomic and molecular processes in the atmosphere.

\section{\label{Sec:Ag and Setup}The Silver D1 line}

\begin{figure}
\includegraphics*[bb= 0 35 500 310, width=3.3in, scale=1]{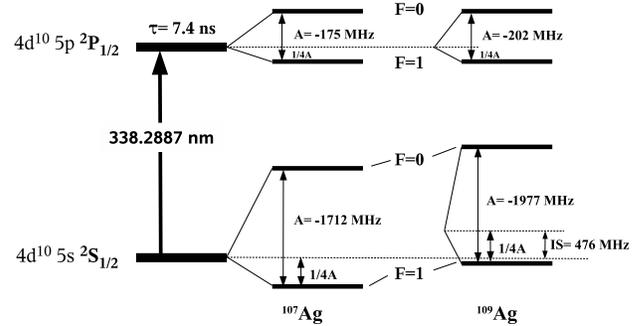}
\caption{\label{fig:AgHF} Hyperfine structure of the Ag D1 line. IS indicates the isotope shift between the two Ag isotopes.}
\end{figure}

          Silver belongs to transition metals, and has two stable isotopes $^{109}$Ag and $^{107}$Ag. The electronic configuration is [Kr]$4d^{10}5s\,^{2}\textbf{S}_{1/2}$. The relevant data for the Ag D1 line are summarized in Table.~\ref{tab:AgRef1}. The D1 line is centered at $338.2887$~nm (29552.061~cm$^{-1}$) \cite{AgLambda}. The data for the isotope shift in the ground state have relatively large uncertainty of about $5\%$ \cite{AgMOT} and were measured for the Ag D2 line. The hyperfine interaction causes $^{2}$\textbf{S}$_{1/2}$ state to split into F~=~0 and F~=~1 components and thus both Ag isotopes are direct analogs to hydrogen (J=1/2, I=1/2) and quite similar to alkalis. The D1 line structure is presented in Fig.~\ref{fig:AgHF}.

\begin{table}
\caption{\label{tab:AgRef1}
Relevant spectroscopic data for the Ag D1 line.}
\begin{ruledtabular}
\begin{tabular}{l|c|c|c}

&$^{107}$Ag&$^{109}$Ag&Ref.\\
\hline
Abundance&$51.84\%$&$48.16\%$&\\
Nuclear spin&$1/2$&$1/2$&\\
Isotope shift, MHz&$0$&$-476(10)$&\cite{AgMOT}\\
A,$\,$MHz of 5s$^2$S$_{1/2}$&$-1712.512111(18)$&$-1976.932075(17)$&\cite{AgHFConst1}\\
A,$\,$MHz of 5p$^2$P$_{1/2}$&$-175.4\,(1.7)$&$-201.6\,(2.6)$&\cite{AgExcitedHFS}\\

\end{tabular}
\end{ruledtabular}
\end{table}

\section{Experimental setup and procedure}

\begin{figure}
\includegraphics*[bb= 60 0 315 215, width=3.4in, scale=1.0]{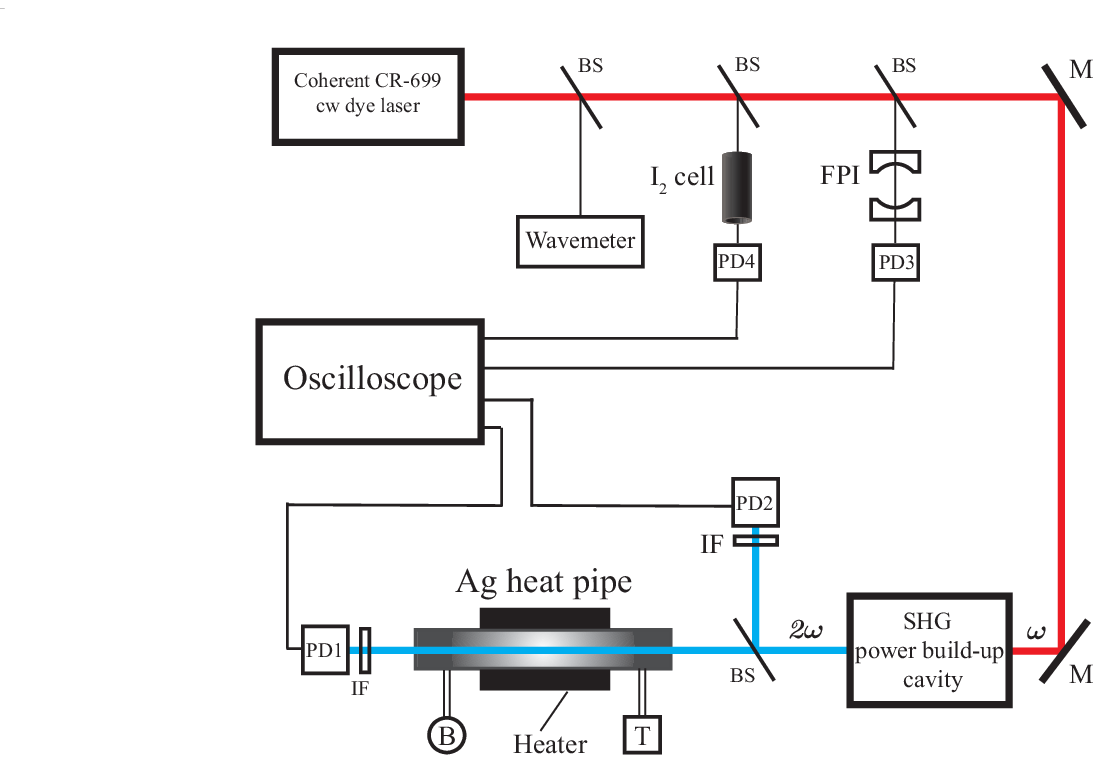}
\caption{\label{fig:ExpSetup}(color online) Experimental setup for measurement pressure broadening and shift of Ag D1 line. BS  beamsplitter; PD - photodiode; I$_{2}$ - iodine absorption cell; FPI - a Fabry-Perot interferometer; SHG - Second Harmonic Generation; IF - interference filters, B - pressure sensor; and T - thermocouple.}
\end{figure}

\begin{figure}
\includegraphics*[width=3.4in]{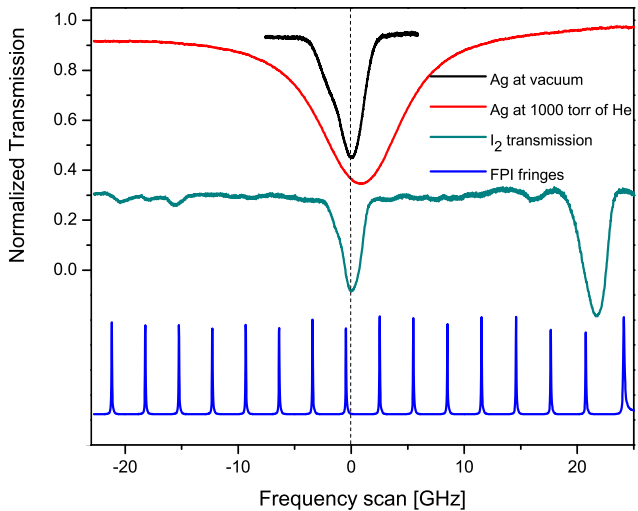}
\caption{\label{fig:TypAbs} Registration of Ag D1 absorption spectra (normalized) for the case of vacuum and a 1000~torr of He. The iodine transmission is shown. The zero of the frequency scale is fixed at the center of an iodine line that is closest to the vacuum Ag absorption. Fabry-Perot interferometer fringes are also shown. Note, that the frequency scale is already multiplied by two, accounting for doubling of the frequency of the dye laser.}
\end{figure}

The setup used for the Ag laser spectroscopy is presented in Fig.~\ref{fig:ExpSetup}. The primary complication for the Ag spectroscopy by cw tunable lasers is that the D1 line lies in the ultraviolet part of the spectrum ($\approx$ 338~nm) and lasers generating this wavelength directly are not available. We use a single-mode ring cavity dye laser (Coherent CR-699) pumped by Ar ion laser (Innova 300) and a modified Microlase MBD-E100 frequency doubler. The dye laser is tuned to a central wavelength of 676.57~nm, and a continuous sweep of the frequency (spanning up to 25~GHz) is performed. The output of the laser is fed into the doubling power build-up cavity, where, via second harmonic generation (SHG) in a LBO (LiB$_{3}$O$_{5}$) crystal, light resonant with the Ag D1 line is produced. The output of the system provides up to $\approx$1~mW of UV light, but for the experiments in this paper the output was reduced to a few microwatt. A Burleigh WA-2000S wavemeter with a resolution of 0.01~nm is used for coarse tuning and monitoring the wavelength of the dye laser. A part of the red light is redirected to a multi-pass iodine absorption cell used as an absolute frequency reference for measurement of the frequency shifts. A confocal Fabry-Perot Interferometer (Tropel~240) with a free spectral range (FSR) of 1.5~GHz is used to obtain frequency markers for calibration and checking the linearity of the laser scan. After the doubler, the UV light is split into two beams. One beam goes to a photodiode (PD2) used as intensity reference. The second beam is directed to a heat-pipe-type absorption cell made of a stainless steel tube supplied with optical windows on both ends. The windows and the two ends of the tube are kept at close to room temperature by active cooling with running water. Heater-wire element situated around the central part of the tube provides the needed temperature of about 1000~K inside the tube where a small piece of silver foil act as an atomic vapor source. A roughing vacuum pump and a manifold for filling the absorption cell with different buffer gases are also attached to the pipe. Buffer-gas pressure is monitored with a calibrated pressure sensor (Baratron 122AA-01000AB) with an accuracy of 1~torr. Temperature sensor is a thermocouple with an accuracy of about 2\%. The light intensity transmitted through the cell is registered with a photodiode (PD1) equipped with an interference filter (IF) with a bandwidth of 10~nm centered at 340~nm.
When scanning the frequency of the dye laser the absorption spectrum of the I$_{2}$ cell in the vicinity of the D1 line is recorded along with the Fabry-Perot fringes and the signals from PD1 and PD2. The Beer's law allows one to present the intensity ratio of the light transmitted through and incident on the vapor as:
\begin{equation}
I(\omega)/I_{0}= \exp[-\sigma_{abs}(\omega)(nL)_{eff}],
\end{equation}
where $\sigma_{abs}$ is the absorption cross-section, $n$ is Ag atomic density and $L$ is the length of the absorption cell. The ``\emph{eff}" subscript indicates that, in heat-pipe type cell, $nL$ is an integral characteristic and is a function of buffer-gas pressure and gradient in the temperature distribution of the atoms along the beam path.
 The absorption cross-section in the case of combined homogeneous (pressure) and inhomogeneous (Doppler) broadening could be represented as a Voigt function with corresponding Lorentzian width $\gamma_{L}$ and Gaussian width $\gamma_{G}$. The normalized measured spectrum is fit with an exponent of the sum of six Voigt functions corresponding to the allowed optical transitions. The best fit Voigt function is calculated using the complex error function (\texttt{wofz}) implemented in the SciPy package. A term linear with the frequency is also added to the fit accounting for a slope in the data which arise, for example, due to interference effect from the windows of the absorption cell. The zero value on the frequency axis is fixed as a center of an I$_{2}$ absorption peak that is closest to the vacuum Ag D1 line center. We note the remarkable coincidence between the center frequencies of the iodine resonance and half the frequency of the silver resonance (see Fig.~\ref{fig:TypAbs}), which could be useful for the frequency-locking purposes.

\begin{table}
\caption{\label{tab:Results}
Broadening and shift parameters for the Ag D1 line by He, Ar, and N$_{2}$ in MHz/torr at 1000~K for the results from this work, and at 1220 K for data from Ref.~\cite{AgBShChen}.}
%\begin{ruledtabular}
\begin{tabular}{|c|c|c|c|c|}
\hline
Buffer Gas&\multicolumn{2}{c|}{Broadening}&\multicolumn{2}{c|}{Shift}\\
\cline{2-5}
&This work&Ref.~\cite{AgBShChen}&This work&Ref.~\cite{AgBShChen}\\
\hline
He&$5.8(1)$&4.1&$+1.16(2)$&+1.15\\
Ar&$5.2(2)$&$4.12$&$-2.28(4)$&-2.54\\
N$_{2}$&$5.2(1)$&--&$-2.52(7)$&--\\
\hline
\end{tabular}
%\end{ruledtabular}
\end{table}

\begin{figure}[b]
\begin{tabular}{ccc}
\centerline{\includegraphics*[width=3.0in]{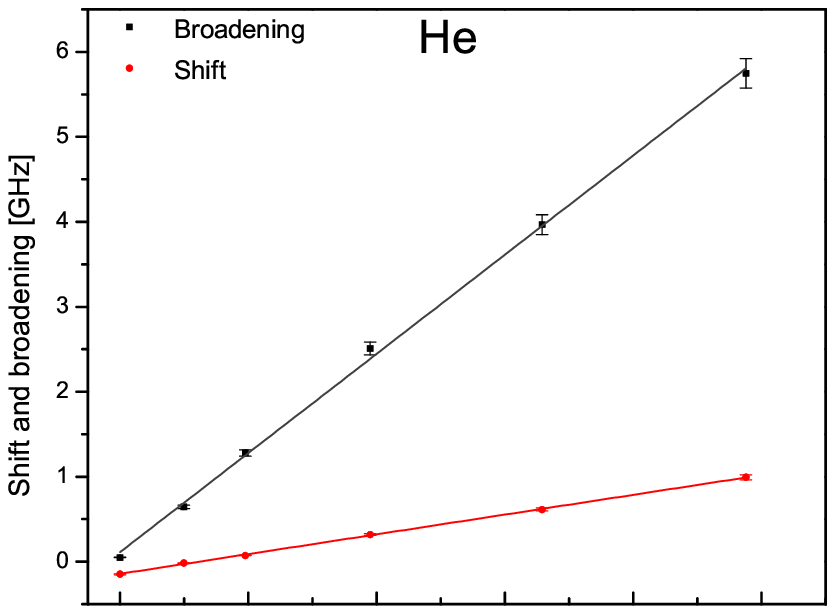}} \\
\centerline{\includegraphics*[width=3.0in]{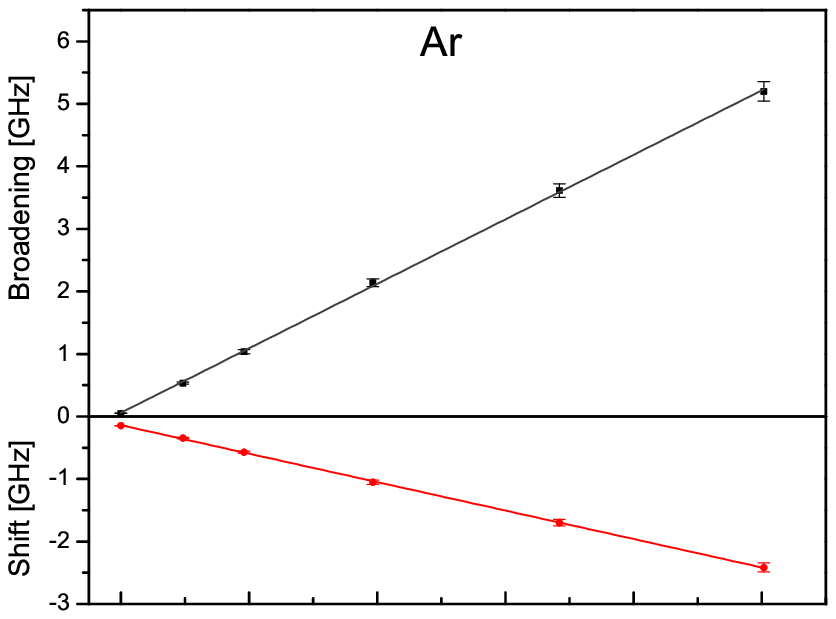}} \\
\centerline{\includegraphics*[width=3.0in]{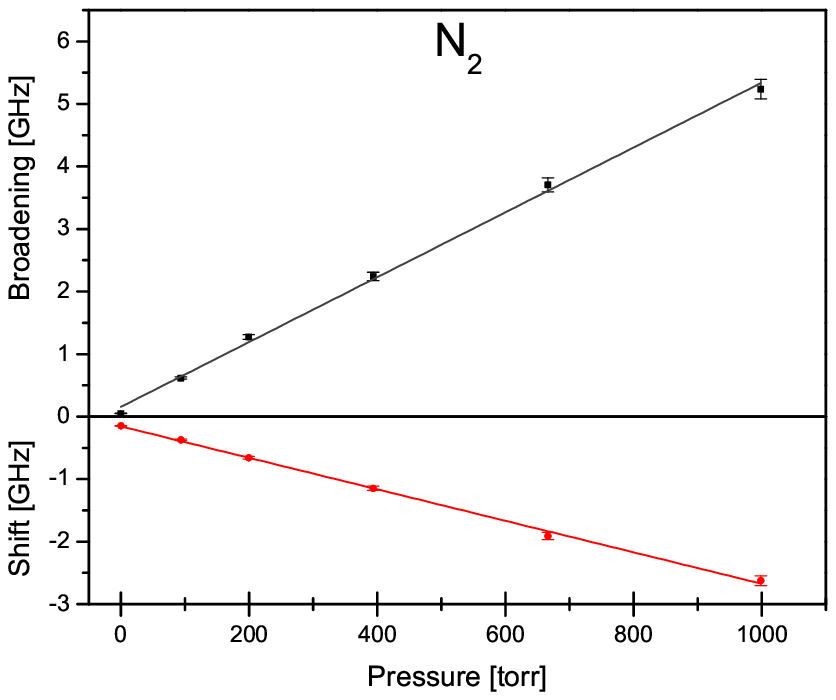}} \\
\end{tabular}
\caption{\label{fig:AllFits}Fits to the experimental data for pressure broadening and shift of the Ag D1 line by He, Ar, and N$_{2}$ at $\approx$1000~K. The error indicated for each point is determined from the best fit to the normalized transmission spectra.}
\end{figure}

\section{Results and Discussion.}
 An illustration of a typical absorption spectrum is presented in Fig.~\ref{fig:TypAbs} for the cases of vacuum (less than $10^{-2}$~torr) and a 1000~torr of He. FPI fringes and I$_{2}$ absorption are also shown. The zero of the frequency scale is set at the center of I$_{2}$ absorption peak that is closest to the vacuum Ag line. The broadening and shift of the resonances are clearly seen.
Fits to all data are presented in Fig.~\ref{fig:AllFits}.
The results for the shift and broadening for He, Ar and N$_{2}$ are presented in Table~\ref{tab:Results} along with the previous measurements \cite{AgBShChen}. All the data are taken in the temperature interval of 990--1030~K.
The errors for the broadening data shown in Table~\ref{tab:Results} are calculated as a sum of the errors from the linear fits shown in Fig.~\ref{fig:AllFits}, the frequency calibration error of about 1.2\%, and the pressure error of about 0.1\%. In determination of the shift errors, an additional error of about 1\% is added accounting for the accuracy in determination of the iodine line center used for frequency reference.
One factor not taken into account in the error budget is the fact that along the laser propagation path the atomic density (Ag and buffer gas) and temperature profile are not flat. However, we benefit from the strong filtering action of the Ag absorption (density) profile itself. The registered absorption is from an ensemble of Ag atoms (in thermal equilibrium with the buffer gas atoms) with a narrow distribution of no more than 50~K around the maximum temperature in the tube. This is due to the fact that the absorption has exponential dependence on the Ag density, which, in turn, depends nearly exponentially on the temperature.
    The results for the shift for He and Ar from \cite{AgBShChen} are in a relatively good agreement with our measured values. However, that is not the case for the broadening, where the discrepancy is about 20\% for Ar and 30\% for He with respect to our results. This comparison is done without taking into account the difference in the temperatures at which the measurements were performed. In fact, if we assume that the collisional cross section for broadening changes negligibly in the temperature interval 1000-1250~K, then the broadening should be proportional only to the relative velocity between Ag atom and the buffer gas atom. In such a case we expect the pressure broadening coefficient to be proportional to the $T^{1/2}$. The latter would further deepen the difference between the results of the two works by an additional 10\% for He and Ar.

\section{Conclussion}We have measured broadening and shift of the Ag D1 line caused by He, Ar, and N$_{2}$ buffer gases. To the best of our knowledge, the data for N$_{2}$ are provided for the first time. The results for the shift for He and Ar are in relatively good agreement with the data from Ref.~\cite{AgBShChen}, but the broadenings are smaller in the earlier work. The values determined in Ref.\cite{AgBShChen} are measured at about 1200~K and at pressures ranging from about 5~atm up to 80~atm. No errors for the shift and broadening are indicated by the authors. In contrast, in the current work we covered the low-pressure range up to 1000~torr and temperature of about 1000~K.
Silver serve as a good proxy for alkali atoms with a big advantage of being nonreactive with, for example, CO$_{2}$ and O$_{2}$. In fact,  in a recent work \cite{Jerome}, it is shown that the total(elastic and inelastic) scattering cross sections for Ag-N$_{2}$ and Na-N$_{2}$ are similar, despite the fact that the inelastic collisions are more important for Ag-N$_{2}$ than for Na-N$_{2}$. We plan to extend our work by measuring broadening and shift due to O$_{2}$, as well as spin depolarization cross-section Ag--O$_{2}$ which will serve as a good proxy for the Na--O$_{2}$ cross section. The latter is important for modeling and realization of highly effective guide stars widely used in most advanced telescopes \cite{SodiumGS}.

\begin{acknowledgments}
The authors would like to thank Brian Patton for numerous suggestions on the manuscript and Jerome Loreau for sharing his results on Ag-N$_{2}$ and Na-N$_{2}$ cross sections. This research has been supported by NSF, grant PHY~-~0855552 and by the Director, Office of Science, Nuclear Science Division of the U.S. Department of Energy under Contract No. DE-AC03-76SF00098.
\end{acknowledgments}

\bibliography{SilverBibl}

%merlin.mbs 2010-03-15 4.21a (PWD, AO, DPC)
%Control: key (0)
%Control: author (8) initials jnrlst
%Control: editor formatted (1) identically to author
%Control: production of article title (-1) disabled
%Control: page (0) single
%Control: year (1) truncated
%Control: production of eprint (0) enabled
\providecommand{\noopsort}[1]{}\providecommand{\singleletter}[1]{#1}%
\begin{thebibliography}{12}%
\makeatletter
\providecommand \@ifxundefined [1]{%
 \@ifx{#1\undefined}
}%
\providecommand \@ifnum [1]{%
 \ifnum #1\expandafter \@firstoftwo
 \else \expandafter \@secondoftwo
 \fi
}%
\providecommand \@ifx [1]{%
 \ifx #1\expandafter \@firstoftwo
 \else \expandafter \@secondoftwo
 \fi
}%
\providecommand \natexlab [1]{#1}%
\providecommand \enquote  [1]{``#1''}%
\providecommand \bibnamefont  [1]{#1}%
\providecommand \bibfnamefont [1]{#1}%
\providecommand \citenamefont [1]{#1}%
\providecommand \href@noop [0]{\@secondoftwo}%
\providecommand \href [0]{\begingroup \@sanitize@url \@href}%
\providecommand \@href[1]{\@@startlink{#1}\@@href}%
\providecommand \@@href[1]{\endgroup#1\@@endlink}%
\providecommand \@sanitize@url [0]{\catcode `\\12\catcode `\$12\catcode
  `\&12\catcode `\#12\catcode `\^12\catcode `\_12\catcode `\%12\relax}%
\providecommand \@@startlink[1]{}%
\providecommand \@@endlink[0]{}%
\providecommand \url  [0]{\begingroup\@sanitize@url \@url }%
\providecommand \@url [1]{\endgroup\@href {#1}{\urlprefix }}%
\providecommand \urlprefix  [0]{URL }%
\providecommand \Eprint [0]{\href }%
\@ifxundefined \urlstyle {%
  \providecommand \doi  [0]{\begingroup \@sanitize@url \@doi}%
  \providecommand \@doi [1]{\endgroup \@@startlink {\doibase
  #1}doi:\discretionary {}{}{}#1\@@endlink }%
}{%
  \providecommand \doi  [0]{doi:\discretionary{}{}{}\begingroup
  \urlstyle{rm}\Url }%
}%
\providecommand \doibase [0]{http://dx.doi.org/}%
\providecommand \Doi [0]{\begingroup \@sanitize@url \@Doi }%
\providecommand \@Doi  [1]{\endgroup\@@startlink{\doibase#1}\@@Doi}%
\providecommand \@@Doi [1]{#1\@@endlink}%
\providecommand \selectlanguage [0]{\@gobble}%
\providecommand \bibinfo  [0]{\@secondoftwo}%
\providecommand \bibfield  [0]{\@secondoftwo}%
\providecommand \translation [1]{[#1]}%
\providecommand \BibitemOpen [0]{}%
\providecommand \bibitemStop [0]{}%
\providecommand \bibitemNoStop [0]{.\EOS\space}%
\providecommand \EOS [0]{\spacefactor3000\relax}%
\providecommand \BibitemShut  [1]{\csname bibitem#1\endcsname}%
%</preamble>
\bibitem [{\citenamefont {Bender}\ \emph {et~al.}(1976)\citenamefont {Bender},
  \citenamefont {Hall}, \citenamefont {Garstang}, \citenamefont {Pichanick},
  \citenamefont {Smith}, \citenamefont {Barger},\ and\ \citenamefont
  {West}}]{AgClock1}%
  \BibitemOpen
  \bibfield  {author} {\bibinfo {author} {\bibfnamefont {P.~L.}\ \bibnamefont
  {Bender}}, \bibinfo {author} {\bibfnamefont {J.~L.}\ \bibnamefont {Hall}},
  \bibinfo {author} {\bibfnamefont {R.~H.}\ \bibnamefont {Garstang}}, \bibinfo
  {author} {\bibfnamefont {R.~M.~J.}\ \bibnamefont {Pichanick}}, \bibinfo
  {author} {\bibfnamefont {W.~W.}\ \bibnamefont {Smith}}, \bibinfo {author}
  {\bibfnamefont {R.~L.}\ \bibnamefont {Barger}}, \ and\ \bibinfo {author}
  {\bibfnamefont {J.~B.}\ \bibnamefont {West}},\ }\href@noop {} {\bibfield
  {journal} {\bibinfo  {journal} {Bull. Am. Phys. Soc},\ }\textbf {\bibinfo
  {volume} {21}},\ \bibinfo {pages} {599} (\bibinfo {year} {1976})}\BibitemShut
  {NoStop}%
\bibitem [{\citenamefont {Badr}\ \emph {et~al.}(2001)\citenamefont {Badr},
  \citenamefont {S.~Gu\'{e}randel},\ and\ \citenamefont {Himbert}}]{AgClock2}%
  \BibitemOpen
  \bibfield  {author} {\bibinfo {author} {\bibfnamefont {T.}~\bibnamefont
  {Badr}}, \bibinfo {author} {\bibfnamefont {P.~J.}\ \bibnamefont
  {S.~Gu\'{e}randel}, \bibfnamefont {M.D.~Plimmer}}, \ and\ \bibinfo {author}
  {\bibfnamefont {M.}~\bibnamefont {Himbert}},\ }\href@noop {} {\bibfield
  {journal} {\bibinfo  {journal} {Eur. Phys. J. D},\ }\textbf {\bibinfo
  {volume} {14}},\ \bibinfo {pages} {39} (\bibinfo {year} {2001})}\BibitemShut
  {NoStop}%
\bibitem [{\citenamefont {Brahms}\ \emph {et~al.}(1997)\citenamefont {Brahms},
  \citenamefont {Newman}, \citenamefont {Johnson}, \citenamefont {Greytak},
  \citenamefont {Kleppner},\ and\ \citenamefont {Doyle}}]{AgExcip}%
  \BibitemOpen
  \bibfield  {author} {\bibinfo {author} {\bibfnamefont {N.}~\bibnamefont
  {Brahms}}, \bibinfo {author} {\bibfnamefont {B.}~\bibnamefont {Newman}},
  \bibinfo {author} {\bibfnamefont {C.}~\bibnamefont {Johnson}}, \bibinfo
  {author} {\bibfnamefont {T.}~\bibnamefont {Greytak}}, \bibinfo {author}
  {\bibfnamefont {D.}~\bibnamefont {Kleppner}}, \ and\ \bibinfo {author}
  {\bibfnamefont {J.}~\bibnamefont {Doyle}},\ }\href@noop {} {\bibfield
  {journal} {\bibinfo  {journal} {Phys.\ Rev. Lett.},\ }\textbf {\bibinfo
  {volume} {79}},\ \bibinfo {pages} {629} (\bibinfo {year} {1997})}\BibitemShut
  {NoStop}%
\bibitem [{\citenamefont {Uhlenberg}\ \emph {et~al.}(2000)\citenamefont
  {Uhlenberg}, \citenamefont {Dirscherl},\ and\ \citenamefont
  {Walther}}]{AgMOT}%
  \BibitemOpen
  \bibfield  {author} {\bibinfo {author} {\bibfnamefont {G.}~\bibnamefont
  {Uhlenberg}}, \bibinfo {author} {\bibfnamefont {J.}~\bibnamefont
  {Dirscherl}}, \ and\ \bibinfo {author} {\bibfnamefont {H.}~\bibnamefont
  {Walther}},\ }\href@noop {} {\bibfield  {journal} {\bibinfo  {journal}
  {Phys.\ Rev. A},\ }\textbf {\bibinfo {volume} {62}},\ \bibinfo {pages}
  {063404} (\bibinfo {year} {2000})}\BibitemShut {NoStop}%
\bibitem [{\citenamefont {Brahms}\ \emph {et~al.}(2008)\citenamefont {Brahms},
  \citenamefont {Newman}, \citenamefont {Johnson}, \citenamefont {Greytak},
  \citenamefont {Kleppner},\ and\ \citenamefont {Doyle}}]{AgMagT}%
  \BibitemOpen
  \bibfield  {author} {\bibinfo {author} {\bibfnamefont {N.}~\bibnamefont
  {Brahms}}, \bibinfo {author} {\bibfnamefont {B.}~\bibnamefont {Newman}},
  \bibinfo {author} {\bibfnamefont {C.}~\bibnamefont {Johnson}}, \bibinfo
  {author} {\bibfnamefont {T.}~\bibnamefont {Greytak}}, \bibinfo {author}
  {\bibfnamefont {D.}~\bibnamefont {Kleppner}}, \ and\ \bibinfo {author}
  {\bibfnamefont {J.}~\bibnamefont {Doyle}},\ }\href@noop {} {\bibfield
  {journal} {\bibinfo  {journal} {Phys.\ Rev. Lett.},\ }\textbf {\bibinfo
  {volume} {101}},\ \bibinfo {pages} {103002} (\bibinfo {year}
  {2008})}\BibitemShut {NoStop}%
\bibitem [{\citenamefont {Clayton}\ and\ \citenamefont
  {Ch'en}(1952)}]{AgBShChen}%
  \BibitemOpen
  \bibfield  {author} {\bibinfo {author} {\bibfnamefont {E.~D.}\ \bibnamefont
  {Clayton}}\ and\ \bibinfo {author} {\bibfnamefont {S.}~\bibnamefont
  {Ch'en}},\ }\href@noop {} {\bibfield  {journal} {\bibinfo  {journal} {Phys.\
  Rev.},\ }\textbf {\bibinfo {volume} {85}},\ \bibinfo {pages} {68} (\bibinfo
  {year} {1952})}\BibitemShut {NoStop}%
\bibitem [{\citenamefont {Sushkov}\ and\ \citenamefont
  {Budker}(2008)}]{AgCryoLife}%
  \BibitemOpen
  \bibfield  {author} {\bibinfo {author} {\bibfnamefont {A.~O.}\ \bibnamefont
  {Sushkov}}\ and\ \bibinfo {author} {\bibfnamefont {D.}~\bibnamefont
  {Budker}},\ }\href@noop {} {\bibfield  {journal} {\bibinfo  {journal} {Phys.\
  Rev. A},\ }\textbf {\bibinfo {volume} {77}},\ \bibinfo {pages} {042707}
  (\bibinfo {year} {2008})}\BibitemShut {NoStop}%
\bibitem [{\citenamefont {Holzloehner}\ \emph {et~al.}(2010)\citenamefont
  {Holzloehner}, \citenamefont {Rochester}, \citenamefont {Calia},
  \citenamefont {Budker}, \citenamefont {Higbie},\ and\ \citenamefont
  {Hackenberg}}]{SodiumGS}%
  \BibitemOpen
  \bibfield  {author} {\bibinfo {author} {\bibfnamefont {R.}~\bibnamefont
  {Holzloehner}}, \bibinfo {author} {\bibfnamefont {S.~M.}\ \bibnamefont
  {Rochester}}, \bibinfo {author} {\bibfnamefont {D.~B.}\ \bibnamefont
  {Calia}}, \bibinfo {author} {\bibfnamefont {D.}~\bibnamefont {Budker}},
  \bibinfo {author} {\bibfnamefont {J.~M.}\ \bibnamefont {Higbie}}, \ and\
  \bibinfo {author} {\bibfnamefont {W.}~\bibnamefont {Hackenberg}},\
  }\href@noop {} {\bibfield  {journal} {\bibinfo  {journal} {AA},\ }\textbf
  {\bibinfo {volume} {510}},\ \bibinfo {pages} {14} (\bibinfo {year}
  {2010})}\BibitemShut {NoStop}%
\bibitem [{\citenamefont {Pickering}\ and\ \citenamefont
  {Zilio}(2001)}]{AgLambda}%
  \BibitemOpen
  \bibfield  {author} {\bibinfo {author} {\bibfnamefont {J.~C.}\ \bibnamefont
  {Pickering}}\ and\ \bibinfo {author} {\bibfnamefont {V.}~\bibnamefont
  {Zilio}},\ }\href@noop {} {\bibfield  {journal} {\bibinfo  {journal} {Eur.
  Phys. J. D},\ }\textbf {\bibinfo {volume} {13}},\ \bibinfo {pages} {181}
  (\bibinfo {year} {2001})}\BibitemShut {NoStop}%
\bibitem [{\citenamefont {Dahmen}\ and\ \citenamefont
  {Penselin}(1967)}]{AgHFConst1}%
  \BibitemOpen
  \bibfield  {author} {\bibinfo {author} {\bibfnamefont {H.}~\bibnamefont
  {Dahmen}}\ and\ \bibinfo {author} {\bibfnamefont {S.}~\bibnamefont
  {Penselin}},\ }\href@noop {} {\bibfield  {journal} {\bibinfo  {journal} {Z.\
  Phys},\ }\textbf {\bibinfo {volume} {200}},\ \bibinfo {pages} {456} (\bibinfo
  {year} {1967})}\BibitemShut {NoStop}%
\bibitem [{\citenamefont {Carlsson}\ \emph {et~al.}(1990)\citenamefont
  {Carlsson}, \citenamefont {Jonsson},\ and\ \citenamefont
  {Sturesson}}]{AgExcitedHFS}%
  \BibitemOpen
  \bibfield  {author} {\bibinfo {author} {\bibfnamefont {J.}~\bibnamefont
  {Carlsson}}, \bibinfo {author} {\bibfnamefont {P.}~\bibnamefont {Jonsson}}, \
  and\ \bibinfo {author} {\bibfnamefont {L.}~\bibnamefont {Sturesson}},\
  }\href@noop {} {\bibfield  {journal} {\bibinfo  {journal} {Z.\ Phys. D},\
  }\textbf {\bibinfo {volume} {16}},\ \bibinfo {pages} {87} (\bibinfo {year}
  {1990})}\BibitemShut {NoStop}%
\bibitem [{\citenamefont {Loreau}()}]{Jerome}%
  \BibitemOpen
  \bibfield  {author} {\bibinfo {author} {\bibfnamefont {J.}~\bibnamefont
  {Loreau}},\ }\href@noop {} {}\bibinfo {note} {Private
  communication}\BibitemShut {NoStop}%
\end{thebibliography}%

\end{document}